# Distribution of atomic rearrangement vectors in a metallic glass

Ajay Annamareddy,[1,*] Bu Wang,[1,2] Paul M. Voyles,[1] and Dane Morgan[1,*]

[1]*Department of Materials Science and Engineering, University of Wisconsin-Madison, Madison, Wisconsin 53706, USA*

[2]*Department of Civil and Environmental Engineering, University of Wisconsin-Madison, Madison, Wisconsin 53706, USA*

[*] Authors to whom correspondence should be addressed: ajeykrishna@gmail.com and ddmorgan@wisc.edu

## Abstract

Short-timescale atomic rearrangements are fundamental to the kinetics of glasses and frequently dominated by one atom moving significantly (a rearrangement), while others relax only modestly. The rates and directions of such rearrangements (or hops) are dominated by the distributions of activation barriers ($E_{act}$) for rearrangement for a single atom and how those distributions vary across the atoms in the system. We have used molecular dynamics simulations of $Cu_{50}Zr_{50}$ metallic glass below $T_g$ in an isoconfigurational ensemble to catalog the ensemble of rearrangements from thousands of sites. The majority of atoms are strongly caged by their neighbors, but a tiny fraction has a very high propensity for rearrangement, which leads to a power-law variation in the cage-breaking probability for the atoms in the model. In addition, atoms generally have multiple accessible rearrangement vectors, each with its own $E_{act}$. However, atoms with lower $E_{act}$ (or higher rearrangement rates) generally explored fewer possible rearrangement vectors, as the low $E_{act}$ path is explored far more than others. We discuss how our results influence future modeling efforts to predict the rearrangement vector of a hopping atom.

## I. INTRODUCTION

Glasses are nonequilibrium solids with hierarchical relaxation dynamics, ranging from picoseconds to many years.[1,2] At short timescales, the glassy state evolves via the thermally induced atomic *rearrangements* (or *hops* or *jumps*) that typically occur at a very low rate. The disordered structure of glass leads to multiple barriers for rearrangement for each atom with varying energies. The rearrangement is characterized by a rapid change in the atom's position that is on the scale of atomic size. These elementary excitations are generally termed as *β*-relaxations and they play a critical role in many aspects of a glass, such as stability[3,4] and mechanical performance.[5] The configuration of neighboring atoms (or "the local environment") plays an essential role in determining whether an atom undergoes a rearrangement or not and what type of rearrangement occurs. However, it is not straightforward to relate rearrangement properties to simple intuitive properties of the local environment. Several studies employing classical molecular dynamics (MD) simulations have been able to connect the local structure and dynamics of atoms in model glasses.[6–8] For example, Ding *et al.* have identified specific atomic packing environments, defined in terms of coordination polyhedra, that are conducive for atomic rearrangements.[6] Recently, Yang and Wang have used the structure-related *Shannon entropy*, quantifying the diversity of Voronoi polyhedra that an atom in a unique local atomic packing will explore during relaxation, to connect structure and dynamics.[7] Machine-learning (ML) methods have also contributed to our understanding by connecting the *features* of an atom's local environment to its propensity for rearrangement.[9–15] For example, Liu and co-workers have used the support vector machine method to define *softness* characterizing the local environment of an atom, and it was demonstrated that softness is strongly correlated to short-time atom dynamics.[9–12] Bapst *et al.* have applied graph neural networks to predict the dynamics of atoms over different timescales, starting from the end of the ballistic regime up to the diffusive regime.[13] More recently, Fan and Ma have used convolution neural networks to identify particles that undergo plastic rearrangements when a load is applied.[14] Atomic trajectories from MD simulations, which in principle follow the exact time evolution of the system when paired with an accurate interatomic potential but have limited timescales, are mainly used in the testing of the above ML methods although some studies use experiments on supercooled colloidal liquids.[16]

The above studies developed approaches to predict the propensity of particle rearrangement from its local environment in glass-forming materials. An important next challenge is to be able to establish the actual vector displacements of atoms during a rearrangement from the initial atomic positions. We refer to such a displacement as a rearrangement vector (denoted by RV here, and also often called a hop or jump vector). Determining the dependence of RVs on local environment will give essential insight into mechanisms controlling atomic dynamics and bring the community a step closer to being able to understanding and simulate more complex and longer-time dynamical processes. In this work we seek to understand how the RVs of an atom depend on its local environment as a first step toward a full predictive model. More specifically, we seek to understand *how many RVs correspond to a given local environment and what factors influence the number of RVs?*

We studied the variation in RV of atoms in the glassy state by performing a large number of short-time independent MD simulations from an isoconfigurational ensemble, which means launching the simulations from the same atomic configuration each time but with momenta assigned randomly from the appropriate Maxwell–Boltzmann distribution.[17] These *independent simulations* are akin to *independent observations* of the evolution of a system. Here, we focused on the RV of just the rearranging atom. Due to its disordered structure, the potential energy landscape (PEL) of a glass is irregular and contains a distribution of depths and barrier heights, and the system overcomes a wide distribution of energy barriers ($E_{act}$) during local rearrangement between neighboring sub-basins.[18–20] The atoms themselves encounter multiple barriers for rearrangement along different directions, with different $E_{act}$ and RVs. From independent simulations of the isoconfigurational ensemble, the probability for an atom with a given local environment to escape its cage and the possible RVs for that atom can be examined. As we shall see, RVs can inform us of the multiplicity of the barriers that an atom has surmounted, and the probability of overcoming these individual barriers is related to the magnitude of the barrier, $E_{act}$. Extending this to all the atoms, a better understanding of the PEL of a glass can, thus, be obtained from our work. We find a power-law variation in the number of atoms with different cage-escaping probabilities and that atoms generally have multiple possible RVs, each associated with its own $E_{act}$. However, we find that the lowest rearrangement barrier energy is a key descriptor for determining how many RVs are active. This observation implies that atoms with the fastest rates show almost unique RVs, as the fast path gets explored more often than others. We also discuss how our results could be extended for longer-time simulations or real glasses.

## II. SIMULATION DETAILS

We used empirical MD simulations to study the binary $Cu_{50}Zr_{50}$ in the glassy state. The atomic interactions are modeled by the embedded atom method (EAM) potential, with the parameters developed by Mendelev *et al.*[21] The $Cu_{50}Zr_{50}$ glass structure was generated as follows: a cubic simulation cell containing 16384 atoms was initially equilibrated at 2000 K for 2 ns and then quenched to 1000 K at a rate of 100 K per 6 ns. The system was subsequently cooled at a rate of 100 K per 60 ns, corresponding to a quench rate of $1.66 \times 10^9$ K/s, and the glass transition temperature ($T_g$) was determined to be 700 K.[22] NPT conditions (at zero nominal pressure) were employed during quenching, and periodic boundary conditions were used in all three directions. The temperature and pressure were controlled using the Nosè–Hoover thermostat and barostat, respectively, and a time step of 1.0 fs was used. Simulations were carried out using the open-source LAMMPS software.[23]

All analyses in this work were performed at 640 K, sufficiently below the $T_g$ so that the local environment of an atom changes little due to thermal fluctuations barring an atomic rearrangement. We first generated 100 unique configurations of $Cu_{50}Zr_{50}$ at 640 K using the quenching protocol described above. For each of these configurations, 1000 independent 2 ps-long simulations were performed under NVT conditions with different starting momenta randomly drawn from a Boltzmann distribution associated with 640 K. The short simulations were intended to better ensure that an atom undergoing rearrangement in independent simulations had very similar local environments before the jump. Hence, both the simulation temperature and timescale are chosen such that for an atom that has not rearranged, it is likely that the spatial distribution of surrounding atoms are similar (or, the central atom sees the same distribution of energy barriers for rearrangement) during independent simulations. Inherent (energy-minimized) structures were used to identify rearrangements, and we identify a rearranging atom when it is displaced by at least 1.25 Å after 2 ps. This 1.25 Å atom displacement length is determined by our desire to focus on RVs that lead to diffusive motion. With that goal in mind, the displacement length should clearly be more than typical thermal vibrations, but those are typically no more than a few tenths of Å. However, we found that most rearrangements up to about 1.25 Å did not have activation energies significantly higher than $k_BT$ when explored with elastic band calculations, suggesting they were best understood as approximately athermal relaxation. Such relaxations do occur in the MD quenched glass since so many metastable states are frozen in during the very rapid MD quenches. However, we believe that these relaxations would not lead to diffusive motion and have, therefore, attempted to exclude them from our analysis. We also note that although 1.25 Å is chosen for this work, we have observed that modest changes to the displacement length (e.g., to 1.5 Å) for identifying the rearrangements do not qualitatively change the results of this work. All the results shown here are averaged over the 100 starting configurations.

## III. RESULTS

Given that the $Cu_{50}Zr_{50}$ is below $T_g$ and the MD trajectories are short (2 ps each), most of the atoms (~96%) were not able to break the cages formed by their neighbors and did not rearrange in any of the 1000 independent simulations. Among the rest, a significant portion of them rearranged in only a few simulations. We characterize an atom by the cage-breaking probability, $p_{CB}$, defined as the fraction of simulations in which the atom has escaped the cage and undergone a rearrangement. For atoms that rearrange at least once during 1000 observations, $10^{-3} \leq p_{CB} \leq 1$. Fig. 1A shows the fraction of atoms vs $p_{CB}$, $f(p_{CB})$, for all the atoms in the system, averaged over the 100 starting configurations. A variation of more than four orders is observed in $f$ that closely follows the power-law $f(p) = cp^{-x}$, where $x$ is the power-law exponent and $c$ is the normalization constant, shown by the green dashed line. The inset of Fig. 1A shows the histogram of the total fraction of atoms scaled by the width of the bin when *logarithmic binning* is used, a standard procedure to ascertain power-law behavior when fluctuations in the tail are high.[24,25] The power-law fit, using the least squares technique, agrees very well with the data, with an $R^2$ value of 0.997, and $x$ ~1.66. In general, a power-law distribution has a much heavier tail compared to the more common distributions, like the exponential, and the behavior in Fig. 1A reflects the diverse nature of short-time dynamics possible in glasses, ranging all the way from very hard to move to very easy to move atoms. We have observed that the power-law exponent, $x$ depends strongly on the number of isoconfigurational runs used to create the histogram plot (e.g., $x$ decreases when going from 100 to 1000 isoconfigurational runs) and, therefore, $x =$ 1.66 may not be intrinsic to the material. However, the power-law exponent is related to the distribution of energy barriers, which control the rearrangement rates of atoms.

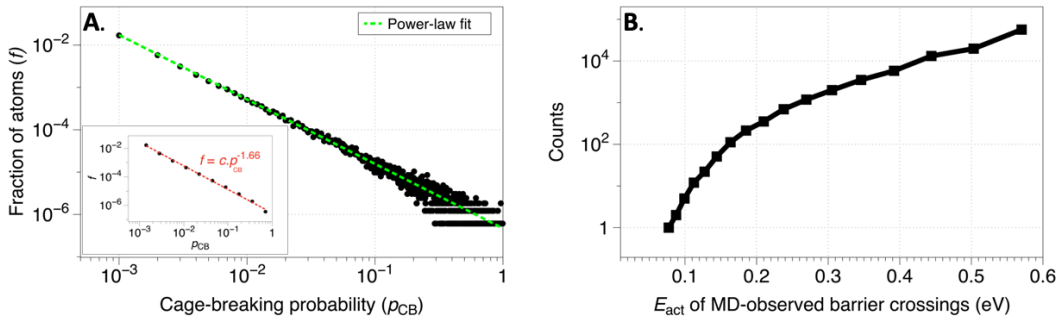

Fig. 1. (A) Variation of the fraction of atoms ($f$) vs the cage-breaking probability ($p_{CB}$) obtained from 1000 independent MD simulations in an isoconfigurational ensemble (with $1/1000 \leq p_{CB} \leq 1$ for rearranging atoms). The dashed line represents the best power-law fit to the data. (inset) Histogram of $f$ vs $p_{CB}$, with bins of equal width on the log scale. The power-law fit is shown as a red dashed line. In general, the power-law exponent (slope of the log-log plot) obtained using logarithmic binning is more accurate than that derived from the original raw data (see Milojevic[25]). (B) Histogram of the activation energies ($E_{act}$) of the observed barrier crossings from the MD simulations.

Atoms in the amorphous state have different barriers for rearrangement along different directions, and a cage-breaking atom may overcome the same barrier or different barriers during rearrangements in independent simulations. The distribution of the rearrangement vectors (RVs) provides information on the distribution of barriers overcome by the atom, and the RVs closely clustered in the same direction are associated with a single barrier. Fig. 2 shows the direction distribution of RVs for an atom as points on the surface of a unit sphere, projected onto the plane of the paper, observed from 1000 independent simulations. The origin or tail of the RVs lie in the center of the sphere, and the RVs are normalized to unit length. An atom in the amorphous state encounters multiple barriers surrounding it, and the different colors of the points on the sphere indicate different barrier crossings. To classify the observed RVs to one or more barriers, we use the following approach. First, we will consider the direction but not the length of the RV in our analysis, as the hop lengths for a given atom are very similar across independent simulations; the average of the standard deviations of all the RV amplitudes is just ~0.12 Å. Thus, the RVs were first normalized (as in Fig. 2). We then classify two RVs to be equivalent (i.e., belong to the same group and correspond to approximately the same activated state) if they were separated by a difference vector with magnitude less than 0.1 (in normalized units, which corresponds to vectors separated by ~5°), and this procedure was repeated until the number of RVs for different activated states stays the same. The value of 0.1 is somewhat arbitrary but matched well with the results of clustering into separate groups of RVs done by human observation on a number of cases. This method has the advantage of being very simple. However, it has the weaknesses that it can incorrectly group quite different RVs if they are part of a connected region. An analysis of all groupings found that the maximum RV separation within a group corresponds to less than 5° in ~99% cases. The atom RVs depicted in Fig. 2 indicate three different energy barriers, and the probability of the atom overcoming the individual barriers ($p_E$), with 1000 observations, is also shown in Fig. 2. The probability $p_{CB}$ (together with the $p_{ES}$ when an atom overcomes separate barriers) can be used to evaluate the barrier-associated rate constant ($k$) using Eqs. (1) or (2) given below (please refer to the Appendix for the derivations).

$$k = -\frac{1}{t}\ln\left(1 - p_{CB}\right), \text{ for a single barrier case} \tag{1}$$

$$k_n = -\frac{1}{t}\frac{\ln\left(1 - p_{CB}\right)}{\left(\frac{\sum_i p_{E,i}}{p_{E,n}}\right)}, \text{ for a multiple-barrier case.} \tag{2}$$

For the multi-barrier case, $k_n$ is the rate constant for the $n^{\text{th}}$ barrier with probability of rearrangement $p_{E,n}$. The rate constant is related to the activation barrier $E_{\text{act}}$ using the equation $k = \nu \exp\left(-E_{\text{act}}/k_B T\right)$, where $\nu$ is the attempt frequency for the process and is generally ~$10^{13}$/s, which is the value used here, $k_B$ is the Boltzmann constant, and $T$ is the absolute temperature. As expected, a low value of $p_E$ (or $p_{CB}$ for single barrier case) leads to a large $E_{\text{act}}$ and vice versa.

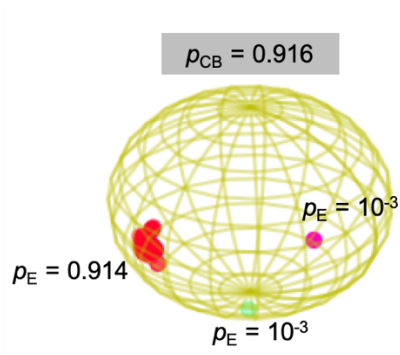

Fig. 2. Illustration of the direction of the rearrangement vectors (RVs) of an atom during 1000 independent simulations as points on the surface of a unit sphere. The rearrangements start at the center of the sphere and are normalized to unit length. An atom in the amorphous state encounters multiple barriers surrounding it, and the different colors of the points on the sphere indicate different barrier crossings, as discerned from the RVs. The values of the overall cage-breaking probability ($p_{CB}$) and the individual barrier-crossing probabilities ($p_E$) are also listed.

The largest energy barrier overcome in our MD simulations, $E_{max}$, can be determined using $p_{CB} = 10^{-3}$ (corresponding to an atom rearranging in just 1 out of our 1000 independent simulations) giving an $E_{max}$ of ~0.55 eV. In experiments on metallic glasses, the typical activation energy of $\beta$-relaxations ($E_\beta$) are around $1 - 2$ eV, and for compositions similar to $Cu_{50}Zr_{50}$, $E_\beta$ values range from $1.6 - 1.8$ eV.[26] In a previous study of $Cu_{50}Zr_{50}$, the nudged elastic band (NEB) method was applied to measure the energy barrier for local rearrangement events from MD simulations to be as high as 1.5 eV although most observed events have much smaller barriers.[27] Our maximum value $E_{max}$ is expected to be lower than that observed in these other contexts due to the limited MD timescales and modest number of independent simulations and because the MD quench structures are a highly unrelaxed class compared to experiments, leading to lower barriers. While $E_{max}$ increases with the number of runs in the isoconfigurational ensemble ($N$), in practice, the increase is bounded as $E_{max}$ varies logarithmically with $N$. With the parameters chosen in this work, $E_{max}$ is ~0.67 eV with $N = 10^4$ runs, and ~0.80 eV with $N = 10^5$ runs.

Fig. 1B shows the histogram of $E_{act}$ of rearranging atoms in our MD simulations obtained from applying Eqs. (1) or (2). The number of activation barriers in the low-energy range is quite small (if there were more, the barriers could easily have been overcome during the simulations), and the concentration increases with increasing $E_{act}$. This result is in qualitative agreement with the investigation of multiple $E_{act}$ for individual atoms in $Cu_{50}Zr_{50}$ by Ding et al. where the number of the observed activation barriers at low energies is quite small and increases with increasing energies for the energy range of Fig. 1B.[19] Ding et al. used the activation–relaxation technique (ART),[28,29] an open-ended saddle point search algorithm, in which an initial perturbation is introduced by randomly displacing (typically by ~0.1 Å) a central atom and a small group of atoms with local connectivity to the central atom. ART then moves the configuration toward a neighboring saddle point, and the activation energy is determined as the energy difference between the saddle point configuration and the initial local energy minimum. Generally, multiple perturbations are introduced for each atom to explore the PEL along different directions, and varying energy barriers may be found. However, the ART method may not find the lowest barriers for a given atom, which the MD simulations do naturally. The present work, therefore, provides a complimentary view of the energy landscape from that in Ding et al.,[19] as this work includes energies more limited in range but focuses on just the RVs and associated barriers that are thermally active.

We now address the question of the distribution of RVs for a given local environment, as observed from independent MD simulations. We quantify the spread ($\Delta$) in RVs from the average of the angles between all possible pairs of RVs associated with the atom,

$$\Delta = \text{average}\left( \sum_{i,j>i} angle(RV_i, RV_j) \right). \quad (3)$$

$\Delta$ is zero for an atom with all its RVs along exactly the same direction and small when one underlying hop dominates with small random variations. To study $\Delta$ for all rearrangement events, we take the following steps:

(i) Divide the rearranging atoms from the MD into bins of varying values of the cage-breaking probability, $p_{CB}$.

(ii) Determine the value $\Delta$ for each atom $i$ in group $p_{CB}$, $\Delta(p_{CB}, i)$.

(iii) Determine the average over all $n$ atoms in the group, $\Delta(p_{CB}) = (1/n)\sum_{i=1}^{n}\Delta(p_{CB}, i)$.

Note that $\Delta$ provides a more sensitive probe of RV distribution than just the number of RV groups since it is sensitive to the relative orientation of RVs (e.g., larger for two RV groups farther apart than two RV groups close together) and the frequency of sampling (larger when distinct RV groups are sampled more evenly vs dominated by one RV group). Fig. 3 shows the variation of $\Delta(p_{CB})$ vs the cage-breaking probability. A monotonic decrease in $\Delta$ is seen with the increase in $p_{CB}$, and $\Delta$ is minimum for the highest $p_{CB}$ atoms. This result implies that the spread in the RV of an atom strongly depends on the barrier distribution surrounding the atom, which determines the cage-breaking probability. In the following, we will show that for the MD-observed rearrangements, the lowest-energy barrier for atoms dominates their cage-breaking probability, i.e., $p_{CB} \approx p_E(E_{min})$ where $E_{min}$ is the activation energy corresponding to the lowest barrier of an atom. Hence, the spread in the RVs of atoms decreases monotonically with a decrease in the lowest-energy barrier of the atoms. Previous studies using *softness*-based ML approaches investigating atomic rearrangements in glassy materials showed that a high propensity for rearrangements is associated with a low enthalpic barrier and also a low rearrangement entropy.[10,30,31] This might suggest that, for atoms with access to small barriers, their hops are accommodated by a small number of nearby sites, which is in agreement with the observations here. It is possible that these atoms have some sort of structural feature (e.g., regions with low atom density) in their vicinity where the rearrangements are mostly directed.

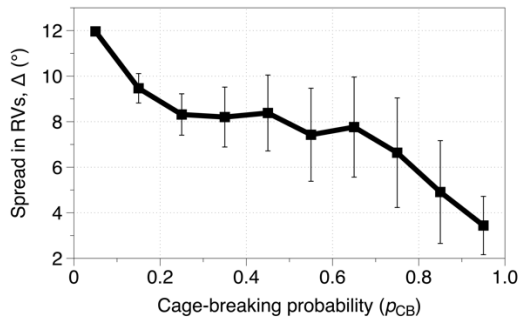

Fig. 3 Variation of the spread ($\Delta$) in the rearrangement vectors against the cage-breaking probability, $p_{CB}$. The standard error of the mean is shown as vertical bars.

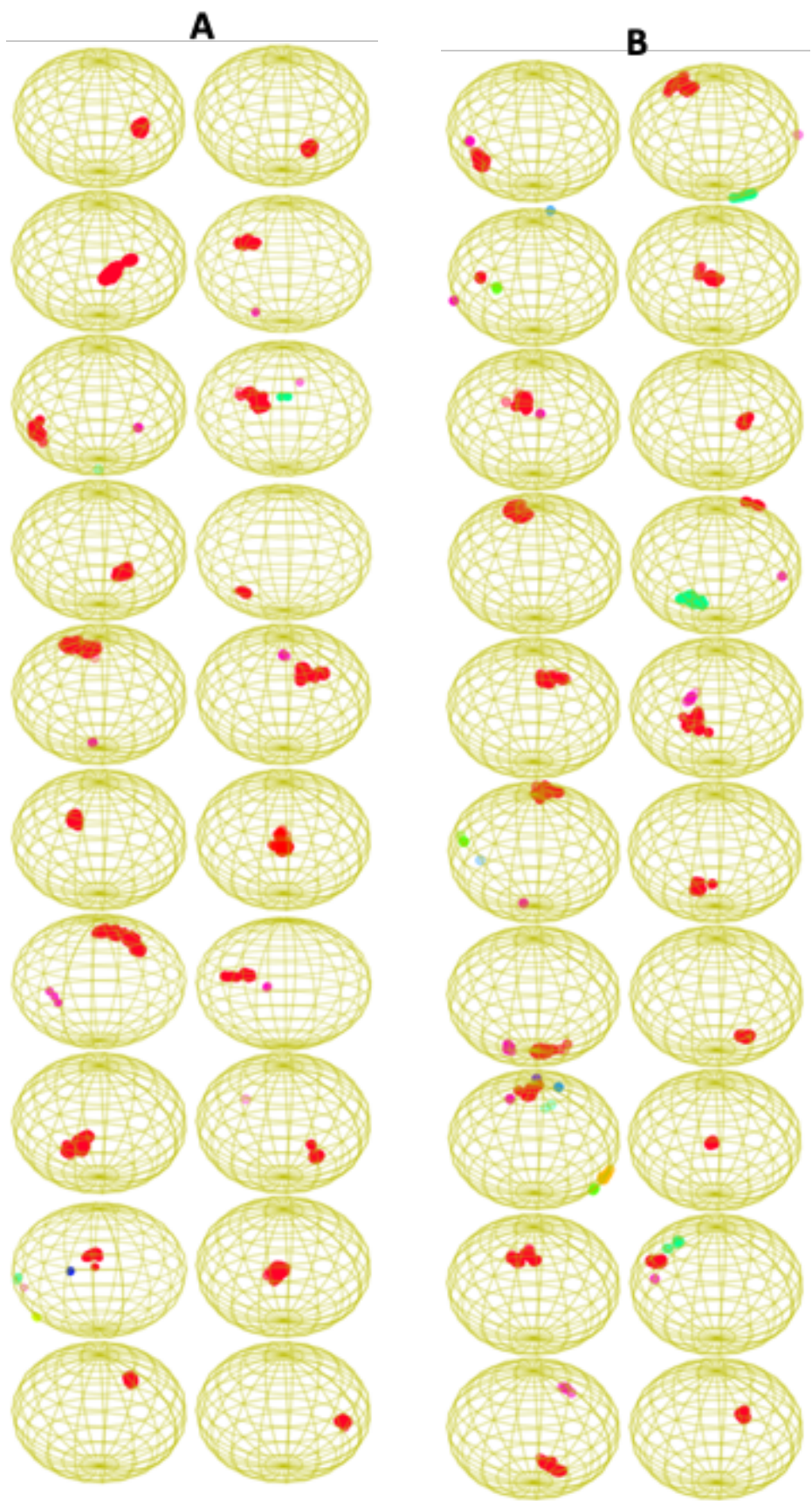

Fig. 4. RV directions for different atoms on a unit sphere with (A) cage-breaking probability $p_{CB} > 0.9$, and (B) $p_{CB}$ ~0.1, from 1000 independent MD simulations. Note that there are more than 900 RVs in (A) and only ~100 in (B).

Fig. 4A shows the RV directions on a unit sphere of all the 20 atoms with $p_{CB} > 0.9$ obtained from our MD simulations. The RVs corresponding to the crossing of different barriers are shown in different colors, with the points in red representing the dominant rearrangement direction (corresponding to the crossing of the lowest barrier seen by the atom). The majority of these atoms have a unique rearrangement direction, although up to five rearrangement directions were observed for one of the atoms. Fig. 4B shows the rearrangement vectors for 20 atoms, picked randomly, with $p_{CB}$ ~0.1. For these atoms, there are more disparate rearrangement directions. The larger number of rearrangement directions and the more uniform exploration of those directions for $p_{CB}$ ~0.1 vs. $p_{CB} > 0.9$ cases are consistent with the trend of $\Delta(p_{CB})$ in Fig. 3.

For the high cage-breaking probability atoms shown in Fig. 4A, even when multiple rearrangement directions are observed, these atoms have a dominant rearrangement direction that accounts for most of the rearrangements. The hops represented in the third row, left panel of Fig. 4A (also shown in Fig. 2) clearly illustrate this case: While there are three unique RV directions for the atom, the dominant rearrangement direction accounts for 99.8% of the total rearrangements. Fig. 5A shows the percentage of rearrangements associated with the dominant rearrangement direction, which quantifies the uniqueness of atomic rearrangements for different cage-breaking probabilities, $p_{CB}$. For the most active atoms with $p_{CB} > 0.9$, almost 100% of the RVs are in the dominant rearrangement direction, and this fraction reduces monotonically with decreasing $p_{CB}$ and reaches ~80% for $p_{CB} = 0.002$ (this is the minimum value of $p_{CB}$ for which multiple rearrangements can be observed with 1000 independent simulations). Fig. 5B illustrates the RVs of the atom for which the dominant rearrangement direction has the lowest contribution (~43%), with $p_{CB}$ at least 0.05, observed from our simulations. In general, atoms have the large majority of their rearrangements taking the same RV, but there is a clear trend for smaller $p_{CB}$ atoms, crossing larger barriers, to have multiple RVs. Therefore, for atoms that overcome larger barriers than observed here, which will occur in longer MD simulations, we expect a multiplicity in the RV of atoms. We delineate two possible extremes with two hypothetical scenarios for the non-uniqueness of RVs for low $p_{CB}$ atoms in Fig. 6. In the first case, shown by the green line, a dominant one, or perhaps just 2–3 RVs still exists, although their role diminishes compared to our MD observations. In the second case, shown by the red dashed line, the RVs are distributed equally in many different directions, and there is a complete breakdown in the uniqueness of the RVs. Our MD results suggest that during long simulations we will see many high barrier rearrangements with multiple RVs. This has important implications for modeling glass evolution, as it implies that multiple atom rearrangements must be modeled for every rearranging atom.

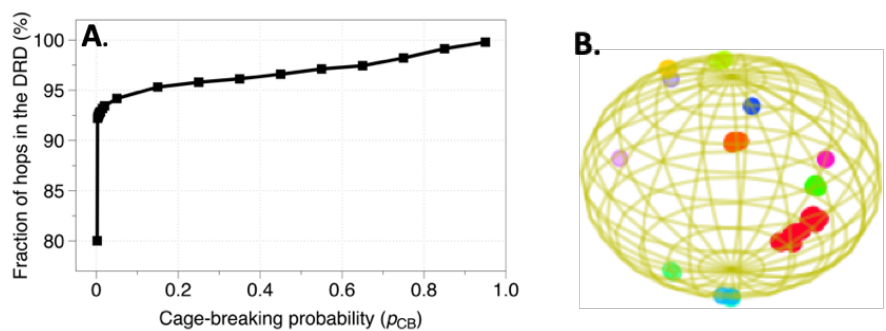

Fig. 5. (A) Quantifying the fraction of rearrangements in the dominant rearrangement direction (DRD) against the cage-breaking probability, $p_{CB}$ of atoms. (B) Illustration of the rearrangement vectors, as points on the surface of a sphere, of an atom for which multiple unique rearrangement directions have been observed, and the fraction of rearrangements in the dominant rearrangement direction for this case is ~43%.

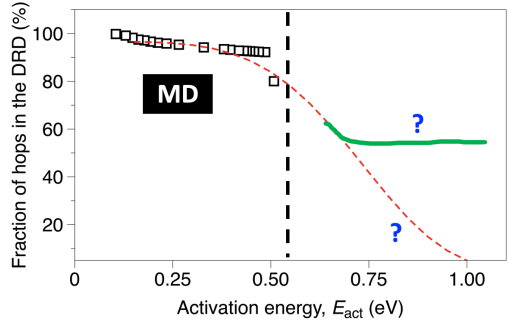

Fig. 6. Extending our MD results, shown by open-square symbols, for the fraction of rearrangements in the dominant rearrangement direction (DRD) towards higher activation energies (beyond the vertical dashed line). We hypothesize that the uniqueness of RVs, as observed for the lowest barriers, vanishes at high activation energies, although the trend is unclear. We depicted two possible scenarios, which are discussed in the text.

## IV. CONCLUSIONS

We have performed molecular dynamics simulations of a model metallic glass to study the nature of short-time rearrangement of atoms in the glassy state. Individual atoms encounter multiple barriers for rearrangement along different directions, and a grand challenge is understanding how the rearrangement vector (RV) of an atom depends on its local environment. To explore this issue, independent MD simulations in an isoconfigurational ensemble were performed at a temperature of interest, and the cage-breaking probabilities of atoms were evaluated based on their dynamical behavior in individual simulations. A power-law variation in the fraction of atoms vs the cage-breaking probability has been observed. From the independent simulations, the variation in the RV of an atom with a given environment can be obtained, and atoms generally have multiple possible rearrangement vectors. However, atoms with lower $E_{act}$ and faster rates generally show fewer possible rearrangement vectors, as the fast path gets explored far more than others. For increasing barrier rearrangements, multiple RVs occur increasingly often. In real glasses and simulations of the order of ns to $\mu$s, the crossing of high barriers by thermal activation becomes quite feasible, so these multiple RV situations are likely to be very important. Therefore, any model for predicting the rearrangement vector of an atom based on the local environment needs to consider the possible multiplicity of rearrangement directions.

## APPENDIX: EXPRESSION RELATING THE BARRIER-CROSSING PROBABILITY AND THE BARRIER ENERGY

When an atom overcomes only a single barrier, the cage-breaking probability $p_{CB}$ can be used to estimate the magnitude of the barrier, $E_{act}$ as follows. The rate constant ($k$) for a barrier crossing with activation energy $E_{act}$ is given by

$$k = \nu \exp(-E_{act}/k_B T),$$ (A1)

where $\nu$ is the attempt frequency for the process and is considered to be ~$10^{13}$/s, $k_B$ is the Boltzmann constant, and $T$ is the absolute temperature. According to rate-theory, the exponential term is interpreted as the probability of a successful crossing. For an atom with a rate constant $k$, such a barrier crossing is a constant rate Poisson process, so the time $t$ for the first cage break has a probability distribution of the form $k \exp(-kt)$,[32] and the probability to overcome the barrier within $t$ is given by

$$p_{CB} = 1 - \exp(-kt) \Rightarrow k = -\frac{1}{t}\ln(1 - p_{CB}).$$ (A2)

Eqs. (A1) and (A2) together give the relationship between $p_{CB}$ and $E_{act}$. As expected, a low value of $p_{CB}$ leads to a large $E_{act}$ and vice-versa. An important caveat to be able to relate the probabilities of rearrangement and the activation energies is that the local environment of the atom is identical in independent observations during rearrangements. The short, 2-ps MD simulations in this work were meant to satisfy this condition, at least approximately.

For the case when an atom overcomes multiple barriers (e.g., as shown in Fig. 2), each individual barrier influences the $p_E$ for all the barriers and all barriers must be taken into account to relate $p_E$ and $E_{act}$. To derive this relation, first we note that the probability distribution for the time of first cage break in the presence of multiple barriers is given by $k_{tot} \exp(-k_{tot}t)$,[32] where the total rate-constant $k_{tot}$ is the sum of individual $k_i$ for different barriers. The probability to overcome any one of the barriers within time $t$ is given by

$$p_{CB} = 1 - \exp(-k_{tot}t) \equiv 1 - \exp\left(-\sum_i k_i t\right). \tag{A3}$$

To evaluate the rate-constant $k_n$ of the $n^{th}$ barrier with the observed probability of rearrangement $p_{E,n}$, we rewrite Eq. (A3) as

$$p_{CB} = 1 - \exp\left(-k_n\left(\frac{\sum_i k_i}{k_n}\right)t\right). \tag{A4}$$

To simplify Eq. (A4), we make use of a well-known relation that for two independent Poisson processes with rate constants $k_1$ and $k_2$, the probability that $s$ events occur in the first process before $t$ events occur in the second process is given by

$$\sum_{j=s}^{s+t-1} \binom{s+t-1}{j} \left(\frac{k_1}{k_1+k_2}\right)^j \left(\frac{k_2}{k_1+k_2}\right)^{s+t-1-j} \tag{A5}$$

where the first term inside the summation represents the binomial coefficient. We are interested in the case of an atom overcoming the $n^{th}$ barrier before all the rest. Without loss of generality, we can treat this case as having two effective processes, $a$ and $b$. Process $a$ is just the process $n$. Process $b$ is all the remaining processes, with the effective rate constant $k_b = \sum_{i \neq n} k_i$. If we take $k_1 = k_a$ and $k_2 = k_b$ in Eq. (A5) and consider $s = t = 1$, it follows that the MD-observed $p_{E,n}$ is equal to $k_n / \sum_i k_i$. This leads to the ratio of any two rate-constants as

$$\frac{k_m}{k_n} = \frac{p_{E,m}}{p_{E,n}}. \tag{A6}$$

Eq. (A6) can be used to simplify Eq. (A4) as

$$p_{CB} = 1 - \exp\left(-k_n\left(\frac{\sum_i p_{E,i}}{p_{E,n}}\right)t\right) \Rightarrow k_n = -\frac{1}{t}\frac{\ln(1-p_{CB})}{\left(\frac{\sum_i p_{E,i}}{p_{E,n}}\right)} \tag{A7}$$

For the case of an atom overcoming a single barrier, the rate-constant from Eq. (A7) simplifies to the rate-constant in Eq. (A2). Again, Eq. (A1) connects $k_n$ and the energy barrier.

## ACKNOWLEDGMENTS

The authors are grateful to the Extreme Science and Engineering Discovery Environment (XSEDE), which is supported by National Science Foundation grant number ACI-1548562, and the Center for High

Throughput Computing (CHTC) at UW-Madison for the computing resources. This work was supported by the University of Wisconsin Materials Research Science and Engineering Center (DMR-1720415).

**DATA AVAILABILITY**

The data that support the findings of this study is made available on figshare at https://doi.org/10.6084/m9.figshare.21377913.v1. All other data relevant to this study is available from AA upon reasonable request.